\documentclass{svmult}

\usepackage{makeidx}         
\usepackage{graphicx}        
\usepackage{multicol}        
\usepackage[bottom]{footmisc}

\makeindex             

\begin{document}

\title*{Analyzing money distributions in `ideal gas' models of markets}
\author{Arnab Chatterjee\inst{1}, Bikas K. Chakrabarti\inst{1}\and
Robin B. Stinchcombe\inst{2}}
\institute{Theoretical Condensed Matter Physics Division and Centre for Applied Mathematics and Computational Science, Saha Institute of Nuclear Physics, Block-AF, Sector-I Bidhannagar, Kolkata-700064, India.
\texttt{arnab@cmp.saha.ernet.in, bikas@cmp.saha.ernet.in}
\and Rudolf Peierls Centre for Theoretical Physics, Oxford University, 1 Keble Road, Oxford, OX1 3NP, UK.
\texttt{stinch@thphys.ox.ac.uk}}
%
%
\maketitle
We analyze an ideal gas like models of a trading market. We propose a new fit 
for the money distribution in the fixed or uniform saving market. For the market
with quenched random saving factors for its agents we show that the steady 
state income ($m$) distribution $P(m)$ in the model has a power law tail 
with Pareto index $\nu$ exactly equal to unity, confirming the earlier 
numerical studies on this model. We analyze the distribution of mutual money
difference and also develop a master equation for the time
development of $P(m)$. Precise solutions are then obtained in some special
cases.

\section{Introduction}
\label{intro}
The distribution of wealth among individuals in an economy has been an
important area of research in economics, for more than a hundred years. 
Pareto \cite{Pareto:1897} first quantified the high-end of the income
distribution in a society and found it to follow a power-law
$P(m) \sim m^{-(1+\nu)}$, where $P$ gives the normalized number of people
with income $m$, and the exponent $\nu$, called the Pareto index, was found
to have a value between 1 and 3.

Considerable investigations with real data during the last ten years revealed
that the tail of the income distribution indeed follows the above mentioned 
behavior and the value of the Pareto index $\nu$ is generally seen to vary 
between 1 and 2.5 \cite{Oliveira:1999,realdatag,realdataln,Sitabhra:2005}. It 
is also known that typically less than $10 \%$ of the population in any country
possesses about $40 \%$ of the total wealth of that country and they follow
the above law. The rest of the low income population, in fact the majority
($90\%$ or more), follow a different distribution which is debated to be either
Gibbs \cite{realdatag,marjit} or log-normal \cite{realdataln}.

Much work has been done recently on models of markets, where economic (trading)
activity is analogous to some scattering process
\cite{marjit,Chakraborti:2000,Chatterjee:2004,Chatterjee:2003,Chakrabarti:2004,othermodels,Slanina:2004}.
We put our attention to models where introducing a saving factor for the
agents, a wealth distribution similar to that in the real economy can be
obtained \cite{Chakraborti:2000,Chatterjee:2004}. Savings do play an important 
role in determining the nature of the wealth distribution in an economy and this
has already been observed in some recent investigations \cite{Willis:2004}.
Two variants of the model have been of recent interest; namely, where the agents
have the same fixed saving factor \cite{Chakraborti:2000}, and where the agents
have a quenched random distribution of saving factors \cite{Chatterjee:2004}.
While the former has been understood to a certain extent (see e.g,
\cite{Das:2003,Patriarca:2004}), and argued to resemble a gamma distribution
\cite{Patriarca:2004}, attempts to analyze the latter model are still 
incomplete (see however, \cite{Repetowicz:2004}). Further numerical studies 
\cite{Ding:2003} of time correlations in the model seem to indicate even more 
intriguing features of the model. 
In this article, we intend to study both the 
market models with savings, analyzing the money difference in the models.

\section{The model}
\label{model}
The market consists of $N$ (fixed) agents, each having money $m_i(t)$ at
time $t$ ($i=1,2,\ldots,N$). The total money $M$ ($=\sum_i^N m_i(t)$) in the
market is also fixed. Each agent $i$ has a saving factor $\lambda_i$
($0 \le \lambda_i < 1$) such
that in any trading (considered as a scattering) the agent saves a fraction
$\lambda_i$ of its money $m_i(t)$ at that time and offers the rest
$(1-\lambda_i)m_i(t)$ for random trading. We assume each trading to be
a two-body (scattering) process. The evolution of money in such a trading
can be written as:
\begin{equation}
\label{mi}
m_i(t+1)=\lambda_i m_i(t) + \epsilon_{ij} \left[(1-\lambda_i)m_i(t) + (1-\lambda_j)m_j(t)\right], 
\end{equation}
\begin{equation}
\label{mj}
m_j(t+1)=\lambda_j m_j(t) + (1-\epsilon_{ij}) \left[(1-\lambda_i)m_i(t) + (1-\lambda_j)m_j(t)\right]
\end{equation}

\noindent
where each $m_i \ge 0$ and $\epsilon_{ij}$ is a random fraction
($0 \le \epsilon \le 1$). 
In the fixed savings market $\lambda_i=\lambda_j$ for all $i$ and $j$, while 
in the distributed savings market $\lambda_i \ne \lambda_j$ with 
$0 \le \lambda_i,\lambda_j < 1$.

\section{Numerical observations} 
\begin{figure}
\centering
\resizebox*{7.5cm}{5.5cm}{\rotatebox{0}{\includegraphics{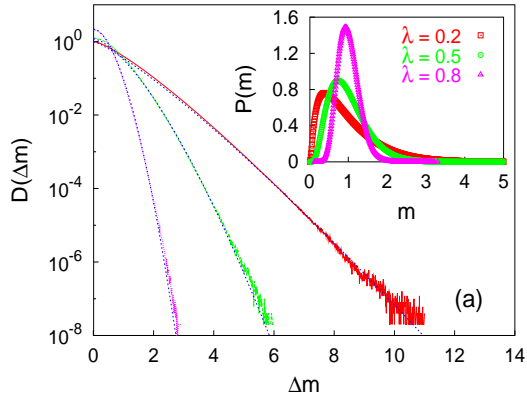}}}
\caption{$D(\Delta)$ in the fixed or uniform
savings market, for $\lambda=0.2,0.5,0.8$ (right to left) and their
fitting curves: $D(\Delta) \sim \exp(-\Delta^{1+\lambda}/T^\prime)$;
the corresponding $P(m)$ the inset.}
\label{fig:1}
\end{figure}

In addition to what have already been reported in Ref.
\cite{Chatterjee:2004,Chatterjee:2003,Chakrabarti:2004} for the model, we
observe that, for the market with fixed or uniform saving factor $\lambda$, 
a fit to Gamma distribution \cite{Patriarca:2004}, 
\begin{equation}
\label{gam}
P(m) \sim m^\eta \exp(-m/T),\quad \eta = \frac{3\lambda}{1-\lambda}
\end{equation}

\noindent
is found to be better than a log-normal distribution. However, our 
observation regarding the distribution $D(\Delta)$ of difference
$\Delta \equiv |\Delta m|$ of money between any two agents in the market
(see Fig. 1a) suggests a different form:
\begin{equation}
\label{new1}
P(m) \sim m^{\delta} \exp(-m^\kappa/T^\prime); \quad \kappa=1+\lambda.
\end{equation}

\noindent
In fact, we have checked, the steady state (numerical) results for $P(m)$
asymptotically fits even better to (\ref{gam}), rather than to (\ref{new1}).

With heterogeneous saving propensity of the agents with fractions $\lambda$
distributed (quenched) widely ($0 \le \lambda < 1$), where the market settles 
to a critical Pareto distribution $P(m) \sim m^{-(1+ \nu)}$ with 
$\nu \simeq 1$ \cite{Chatterjee:2004}, the money difference behaves as 
$D(\Delta m) \sim (\Delta m)^{-(1+ \gamma)}$ with $\gamma \simeq 1$. 
In fact, this behavior is invariant even if we set 
$\epsilon_{ij}=1/2$ \cite{Chatterjee:2005}.
This can be justified by the earlier numerical observation
\cite{Chakraborti:2000,Chatterjee:2004} for fixed $\lambda$ market
($\lambda_i = \lambda$ for
all $i$) that in the steady state, criticality occurs as $\lambda \to 1$
where of course the dynamics becomes extremely slow. In other words,
after the steady state is realized, the third term containing $\epsilon =1/2$
becomes unimportant for the critical behavior.
We therefore concentrate on this case in this paper.

\section{Analysis of money difference} 
\label{moneydiff}
In the process as considered above, the total money $(m_i+m_j)$
of the pair of agents $i$ and $j$ remains
constant, while the difference $\Delta m_{ij}$ evolves for
$\epsilon =1/2$ as
\begin{equation}
\label{dDelmt}
(\Delta m_{ij})_{t+1} = \alpha_{ij}(\Delta m_{ij})_t + \beta_{ij}(m_i+m_j)_t,
\end{equation}

\noindent
where $\alpha_{ij}=\frac{1}{2}(\lambda_i+\lambda_j)$ and
$\beta_{ij}=\frac{1}{2}(\lambda_i-\lambda_j)$. As such, $0 \le \alpha < 1$ and
$-\frac{1}{2} < \beta < \frac{1}{2}$.
The steady state probability distribution $D(\Delta)$ can be written as
(cf. \cite{Chatterjee:2005}):
\begin{eqnarray}
\label{DDel}
D(\Delta)
&=& \int d \Delta^\prime \; D(\Delta^\prime) \;
\langle 
\delta (\Delta -(\alpha + \beta) \Delta^\prime) +
\delta (\Delta -(\alpha - \beta) \Delta^\prime) 
\rangle \nonumber\\
&=& 
2 \langle 
\left( \frac{1}{\lambda} \right)
\; D
\left( \frac{\Delta}{\lambda} \right)
\rangle,
\end{eqnarray}

\noindent
where we have used the symmetry of the $\beta$ distribution and the relation
$\alpha_{ij} + \beta_{ij}=\lambda_i$, and have suppressed labels $i$, $j$.
Here $\langle \ldots \rangle$ denote average over $\lambda$ distribution
in the market.
Taking now a uniform random distribution of the saving factor $\lambda$,
$\rho(\lambda) = 1$ for $0 \le \lambda < 1$, and assuming
$D(\Delta) \sim \Delta^{-(1+\gamma)}$ for large $\Delta$, we get
\begin{equation}
\label{gammaex}
1=2 \int d \lambda \; \lambda^\gamma = 2 (1+\gamma)^{-1},
\end{equation}

\noindent
giving $\gamma=1$.
No other value fits the above equation. This also
indicates that the money distribution $P(m)$ in the market also follows a
similar power law variation, $P(m) \sim m^{-(1+\nu)}$ and $\nu=\gamma$.

\section{Master equation approach}
We also develop a Boltzmann-like master equation for the time
development of $P(m,t)$, the probability distribution of money in the
market \cite{Chatterjee:2005}. We again consider the case 
$\epsilon_{ij}=\frac{1}{2}$ in (\ref{mi}) and (\ref{mj}) and rewrite them as
\begin{equation}
\label{Amat1}
\left( 
\begin{array}{c}
m_i\\m_j
\end{array}
\right)_{t+1}
= \mathcal{A}
\left( 
\begin{array}{c}
m_i\\m_j
\end{array}
\right)_t
{\rm where}\;\;\;
\mathcal{A}=
\left( 
\begin{array}{cc}
\mu_i^+ & \mu_j^-\\
\mu_i^- & \mu_j^+
\end{array}
\right);\quad
\mu^\pm = \frac{1}{2} (1 \pm \lambda).
\end{equation}

\noindent
Collecting the contributions from terms scattering in and subtracting
those scattering out, we can
write the master equation for $P(m,t)$ as
\begin{equation}
\label{partial}
\frac{\partial P(m,t)}{\partial t} + P(m,t) =
\langle
\int d m_i \int d m_j \; P(m_i,t)P(m_j,t)\;
\delta(\mu_i^+ m_i + \mu_j^- m_j -m)
\rangle,
\end{equation}

\noindent
which in the steady state gives
\begin{equation}
\label{aftrpartial}
P(m) =
\langle
\int d m_i \int d m_j \;P(m_i)P(m_j)\;
\delta(\mu_i^+  m_i +\mu_j^- m_j -m)
\rangle.
\end{equation}

\noindent
Assuming, $P(m) \sim m^{-(1+\nu)}$ for
$m \rightarrow \infty$, we get \cite{Chatterjee:2005}
\begin{equation}
\label{mugama}
1
= 
\langle
(\mu^+)^\nu + (\mu^-)^\nu
\rangle
\equiv
\int \int d\mu^+ d\mu^- p(\mu^+) q(\mu^-)
\left[ 
(\mu^+)^\nu + (\mu^-)^\nu
\right].
\end{equation}

\noindent
Considering now the dominant terms 
($\propto x^{-r}$ for $r>0$, or $\propto \ln (1/x)$ for $r=0$) in the
$x \to 0$ limit of the integral 
$\int_0^\infty m^{(\nu + r)} P(m) \exp (-mx) dm$, we 
get from eqn. (\ref{mugama}), after integrations,
$1=2/(\nu + 1)$, giving finally $\nu=1$.

\section{Summary}
We consider the 
ideal-gas-like trading markets where each agent is identified with a gas
molecule and each trading as an elastic or money-conserving (two-body) 
collision 
\cite{Chakraborti:2000,Chatterjee:2004,Chatterjee:2003,Chakrabarti:2004}.
Unlike in a gas, we introduce a saving factor $\lambda$ for each
agents. Our model, without savings ($\lambda=0$), obviously yield a Gibbs
law for the steady-state money distribution. 
Our numerical results for 
uniform saving factor suggests the equilibrium distribution $P(m)$ to be
somewhat different from the Gamma distribution reported earlier 
\cite{Patriarca:2004}.

For widely distributed (quenched) saving factor $\lambda$, numerical studies 
showed \cite{Chatterjee:2004,Chatterjee:2003,Chakrabarti:2004}
that the steady state
income distribution $P(m)$ in the market has a power-law tail
$P(m) \sim m^{-(1+\nu)}$ for large income limit, where $\nu \simeq 1.0$, and
this observation has been confirmed in several later numerical studies as well
\cite{Repetowicz:2004,Ding:2003}. It has been noted from these numerical 
simulation studies that the large income group people usually have larger
saving factors \cite{Chatterjee:2004}. This, in fact, compares well with
observations in real markets \cite{Willis:2004,Dynan:2004}. The time 
correlations induced by the random saving factor also has an interesting
power-law behavior \cite{Ding:2003}. A master equation for 
$P(m,t)$, as in (\ref{partial}), for the original case (eqns. (\ref{mi}) and  
(\ref{mj})) was first formulated for fixed $\lambda$ ($\lambda_i$ same for all
$i$), in \cite{Das:2003} and solved numerically. Later, a generalized master
equation for the same, where $\lambda$ is distributed, was formulated
and solved in \cite{Repetowicz:2004} and \cite{Chatterjee:2005}.
We show here that our analytic study (see \cite{Chatterjee:2005} for details)
clearly support the power-law for $P(m)$ with the exponent value
$\nu=1$ universally, as observed numerically earlier
\cite{Chatterjee:2004,Chatterjee:2003,Chakrabarti:2004}.

\section{Acknowledgments}
BKC is grateful to the INSA-Royal Society Exchange Programme for financial
support to visit the Rudolf Peierls Centre for Theoretical Physics, 
Oxford University, UK and
RBS acknowledges EPSRC support under the grants GR/R83712/01 and GR/M04426
for this work and wishes to thank the Saha Institute of Nuclear Physics
for hospitality during a related visit to Kolkata, India.



\printindex
\end{document}